\documentstyle[12pt]{article}
\title{
Singularity of the Vortex Density of States in $d$-wave Superconductors.}
\author{N.B. Kopnin and G.E. Volovik\\
Low Temperature Laboratory,
Helsinki University of Technology\\
Otakaari 3A, 02150 Espoo, Finland\\
and\\
L.D. Landau Institute for Theoretical Physics, \\
Kosygin Str. 2, 117940 Moscow, Russia\\
}
\begin{document}
\maketitle
\begin{abstract}
{ In $d$-wave superconductors, the electronic density of states (DOS)
induced by a vortex exhibits a divergency at low energies: $N_{vortex}(E)
\sim 1/|E|$. It is the result of  gap nodes in the excitations
spectrum outside the vortex core. The heat capacity in two regimes,
$T^2/T_c^2
\gg B/B_{c2}$ and $T^2/T_c^2 \ll B/B_{c2}$, is discussed.}
\end{abstract}

\section{Introduction}

Two different energy scales, Fermi energy $E_F$ and the gap
amplitude $\Delta \ll E_F$, govern the dynamics of   fermions in
superconductors. This leads to several levels of  description of
excitations in the inhomogeneous background of the order parameter produced
by vortices and textures. (1) A quantum-mechanical approach was used for
calculation of the discrete spectrum of bound states in the vortex core
\cite{Caroli1964}.  This calculations revealed the existence of the
low-energy branch
$$E_0(Q)=-Q\omega_0~~\eqno(1.1)$$
where the orbital quantum number $Q$ is  half of odd integer for
conventional vortices. Here we consider a two-dimensional case, i.e.
we neglect the
dependence on the momentum  $p_z$ along the vortex axis. The interlevel
distance of bound states is small compared to the gap, typically
$\omega_0\sim \Delta^2/ E_F$.

(2) In the quasiclassical (or Eilenberger) approach, quantum mechanics
is applied only for the motion along the trajectory. The energy spectrum
is characterized by the continuous impact parameter $b$ (or the continuous
orbital quantum number $Q=p_Fb$) and by the trajectory direction (angle
$\alpha$) in $x,y$ plane \cite{KramerPesh}. The low-energy part of the
spectrum corresponds to the chiral branch in Eq.(1.1):
$$E_0(Q,\alpha)=-Q\omega_0(\alpha)~~.\eqno(1.2)$$
In a conventional axisymmetric vortex in   $s$-wave superconductors,
$\omega_0$ does not depend on $\alpha$. In a  $d$-wave superconductor, the
gap $\Delta(\alpha)$ is anisotropic, it has the underlined tetragonal
symmetry and exhibits four gap nodes. The typical example which displays
these properties is
$\Delta(\alpha)\propto \cos(2\alpha)$, which has four gap nodes at
$\alpha_0=(2k+1)\pi/4$ with integer $k$. In the vicinity of each gap node
$\Delta(\alpha)\approx \Delta'\vert_{\alpha_0} (\alpha -\alpha_0)$. As a
result the energy spectrum in Eq.(1.2) has a four-fold symmetry and is
given by
\cite{d-waveVortex}
$$ \omega_0(\alpha)\approx (\alpha -\alpha_0)^2~{(\Delta')^2\over E_F}\ln
{1\over
\vert  \alpha -\alpha_0\vert}\eqno(1.3)$$
in the vicinity of the gap nodes.
Properties of the spectrum in the whole range of $E$ and $\alpha$ was
discussed in \cite{SchopohlMaki,Morita} in connection to the scanning
tunneling microscopy experiment \cite{Maggio-Aprile}.

The quantum limit of Eq.(1.1) is restored by using the Bohr-Sommerfeld
quantization rule for the  orbital
momentum $Q$ and the canonically conjugated angle $\alpha$:
$$\int_0^{2\pi} d\alpha~ Q(\alpha,E)=2\pi (n
+\gamma)~~,~~ Q(\alpha,E)=-{E\over \omega_0(\alpha)} ~~,\eqno(1.4)$$
where $n$ is an integer and $\gamma$ is of order unity.
This gives the discrete levels
$$ E=-(n+\gamma) \left[ \int {d\alpha \over 2\pi} {1\over
\omega_0(\alpha)}\right]^{-1} ~~.\eqno(1.5)$$
For axisymmetric vortices in a $s$-wave
superconductor,  Eq.(1.1) with $Q=n+1/2$ is restored if one chooses
$\gamma=1/2$. Actually the choice of $\gamma$  (i.e. either  $\gamma=1/2$ or
$\gamma=0$ ) is dictated by the symmetry of the superconducting state
in presence of the vortex \cite{Misirpashaev1995}.

For the $d$-wave case of Eq.(1.3), the integral in Eq.(1.4) diverges at
angles  close to the gap nodes. This indicates a singular behavior of the
energy spectrum, which originates from $1/r$ tail of the superfluid velocity
$\vec v_s(\vec r)$ far from the vortex core.

(3) At large distances from the core, a pure classical approach can be used
in which the coordinate $\vec r$ and the momentum $\vec p$ of the fermionic
quasiparticles are considered as commuting variables. The energy spectrum is
given by
$$E=\pm \sqrt{\epsilon^2_k + \Delta^2(\alpha)} + \vec k \cdot\vec
v_s(\vec r) ~~,\eqno(1.6)$$
where $\epsilon _k=v_F(k-k_F)$.
This approach has been used for  calculations of a non-analytical behavior
caused by the gap nodes. In  $^3$He-A, the order parameter texture induces
the effective  $\vec v_s(\vec k,\vec r)$ and the point nodes lead  to a
non-analytical density of the normal component at $T=0$
\cite{VolovikMineev1981}. The same effect of  gap nodes resulting in a
nonzero DOS in  presence of the conventional
$\vec v_s$ was discussed in \cite{Muzikar1983,Nagai1984,Xu1995}. Due to
an inhomogeneous vortex-induced velocity $\vec v_s(\vec r)$
in the mixed state
of a $d$-wave superconductor, a  non-analytical dependence of DOS on
magnetic field has been found: $N(0)\sim N_F (B/B_{c2})^{1/2}$, where
$N_F$ is
DOS in the normal state \cite{Volovik1988,d-waveVortex}. The numerical
factor in this dependence has been calculated in \cite{WonMaki}. Possible
experimental realization of such behavior was discussed in
\cite{Moler1994,Fisher1995,Revaz1996}.

Here we consider the energy-dependent  DOS, $N(E)$, for an isolated vortex
and find that $N(E) \propto 1/|E|$ at $E\rightarrow 0$  using both the
quasiclassical and classical approaches.

\section{$1/E$ divergence of DOS in the classical approximation.}

DOS in a homogeneous two-dimensional $d$-wave  superconductor is
$$N(E)=2\int {d^2k\over (2\pi)^2}~\frac{1}{2}\delta
(E\mp \sqrt{\epsilon^2_k + \Delta^2(\alpha)})~~.\eqno(2.1)$$
Here the factor 2 accounts for the spin degrees of freedom and the factor
1/2 takes care of the double counting of particles and holes with
different signs of the square root under the $\delta$-function.

For small $E$, the main contribution comes
from the four gap nodes (see also \cite{Xu1995}):
$$N(E)=2{m\over \pi \Delta'}|E|=2 N_F {|E|\over \Delta'}~~,\eqno(2.2)$$
where $\Delta'=\partial_\alpha\Delta$ at the gap node and  $N_F=m/\pi$ is
the DOS in the normal state. This leads to a $T^2$-dependence of the
specific heat whose experimental evidences are discussed in
\cite{Moler1994,Momono1996}.

The vortex contribution comes mainly  from the superflow around the
vortex and from momenta close to the gap nodes $\vec k_a$
\cite{d-waveVortex}. The locations of the gap nodes are not important,
however we assume the tetragonal symmetry with $\vec k_a=\pm
k_F \hat x,\pm k_F \hat y$ for simplicity.  All four nodes give equal
contributions thus the extra vortex-induced DOS is
$$
N_{d-vortex}(E) =N_F{4\over \Delta'} \int  d^2r~(\vert E-k_F\hat x\cdot\vec
v_s(\vec r)\vert -|E|)~~.\eqno(2.3)$$
The velocity far from the vortex is $\vec v_s(\vec r)=\hat \phi
(\hbar/2mr)$. Using the variable $u=r(2mE/\hbar k_F)$ one has
$$N_{d-vortex}(E)=N_F{v_F^2\over   {\Delta'  |E|}}
\int_0^{2\pi} d\phi \int_0^1  du~(\vert u-\cos\phi\vert
-u)={\pi\over 2}N_F{v_F^2\over
\Delta'  |E|} ~~.\eqno(2.4)$$

The characteristic dimension of the region near the vortex which contributes
to DOS is
$$r(E) ={\hbar k_F\over {2m|E|}} ~~.\eqno(2.5)$$
When $E$ decreases, the size
$r(E)$ reaches the intervortex distance $R_B\sim \xi (B_{c2}/B)^{1/2}$ in
the vortex lattice. For lower energies, $E$ should be substituted with
$(\hbar k_F/2m R_B)$, and the square-root dependence of the DOS in the
lattice of the $d$-wave vortices\cite{d-waveVortex} is restored. So the
Eq.(2.4) holds for the energy $E$ in the range $\Delta' (B/B_{c2})^{1/2} \ll
|E|\ll \Delta'$.

\section{Quasiclassical approximation.}

In this approach, the radial motion is quantized. We consider only  the
chiral energy branch. The two remaining variables are: the impact
parameter
$b$ (or the angular momentum $Q=k_Fb$) and the angle $\alpha$ of the
direction of linear momentum $\vec k$ in the $x,y$ plane.  DOS in this
approximation is
$$N_{d-vortex}(E)=\int {d\alpha~ dQ\over 2\pi}
\delta(E-E_0(Q,\alpha))~~,\eqno(3.1)$$
where $E_0(Q,\alpha) =-Q\omega_0(\alpha)$. Note one sign under the
$\delta$-function as compared to Eq. (2.1): now there is only
one type of excitations on the chiral branch.

For the $d$-wave case where
$\omega_0(\alpha)$ is given by Eq.(1.3), the integral in Eq.(3.1)  diverges
near the nodes: ($\int d\alpha /\omega_0(\alpha)\sim \int d
\alpha / (\alpha-\alpha_0)^2 $, see \cite{d-waveVortex}). This divergence is
related to the large extension of the radial wave function  $\Psi(\vec r,
\alpha)$ when $\alpha$ is close to the direction of the gap nodes.
To treat this divergence properly, one must include  $\Psi(\vec r, \alpha)$
into the equation for DOS explicitly:
$$N_{d-vortex}(E)=\int {d\alpha~ \over 2\pi} \int d^2r
\vert \Psi(\vec r, \alpha)\vert^2
\delta(E- k_F r\sin(\phi-\alpha)\omega_0(\alpha))~~.\eqno(3.2)$$
Here we introduced the impact
parameter $b= r\sin(\phi-\alpha)$. The radial wave function
 at large distances $r\gg \xi$ in a vicinity of the gap node
can be obtained from the Eilenberger equations \cite{Kopnin}:
$$\vert \Psi(\vec r, \alpha)\vert^2= m \Delta'\vert\alpha-\alpha_0\vert
\exp\left(-2r{\Delta'\over v_F}\vert\alpha-\alpha_0\vert
\right)~~.\eqno(3.3)$$
Eqs.(3.2-3) solve the problem of the divergence,
since they have a wider range of applicability than those in
Ref.\cite{d-waveVortex}, where the condition of small impact parameter
compared to the radial extent of the wave function was used. Under
this condition, Eqs.(3.2-3) indeed transform to Eq.(3.1).

We integrate Eq.(3.2) over $\phi$ introducing variables $r=\rho
v_F/2\Delta'$, $\alpha-\alpha_0=\alpha$  and  the parameter
$a=L\Delta'/|E|$ (where $L=-\ln \vert\alpha\vert$). Summing over 4 gap nodes,
one obtains
$$N_{d-vortex}(E)= {2mv_F^2\over {\pi\Delta'|E|}} \int_{0}^{\infty}
d\alpha~\alpha   \int_{0}^{\infty}
\rho d\rho e^{-\rho\alpha}{1\over
\sqrt{(a\rho\alpha^2)^2-1}}~\Theta(a\rho\alpha^2-1) ~~.\eqno(3.4)$$
This integral
is independent of $a$ and is equal to $\pi/2$, thus
$$N_{d-vortex}(E)=
N_F  \pi {v_F^2\over {\Delta'|E|}} ~~.\eqno(3.5)$$
This is two times larger than Eq.(2.4) obtained using classical
considerations. We can thus conclude that the classical approach
is not able to give the correct numerical factor though it feels the
relevant physics and provides a correct order-of-magnitude estimate.
The same is known for DOS in  $^3$He-A textures: the exact DOS obtained
quantum-mechanically in \cite{Dombre} is two times the classical result
of \cite{VolovikMineev1981}.

\section{Conclusion}

The singular behavior of DOS can be seen in the temperature dependence of
the cpecific heat in presence of a vortex lattice. Being averaged over
vortices, DOS (per one  CuO$_2$ superconducting layer)  contains the
bulk term   Eq.(2.2) and the vortex term
$n_L N_{d-vortex}(E) $, where $n_L=B/\Phi_0$ is the flux-line density,
and $\Phi_0$ is the magnetic flux quantum:
$$N(E)= {2 p_F\over \pi v_F \Delta'}|E| + {B\over \Phi_0} {v_Fp_F\over  |E|
\Delta'}~~.
\eqno(4.1)$$
This results in the specific heat:
$$C(T)=\int_{-\infty}^\infty N(E)\, \cosh^{-2}
\left(\frac{E}{2T}\right)~\frac{E^2\, dE}{4T^2}=$$
$$=   18\zeta(3){p_F\over \pi v_F\Delta'} T^2 +
2\ln 2 {B\over\Phi_0}{v_Fp_F\over \Delta'}~~,~~  \sqrt{B\over B_{c2}} \ll
{T\over T_c}\ll 1~~.
\eqno(4.2)$$
The second term in Eq.(4.2) is the temperature-independent linear-in-field
correction of order $ N_F T_c  B/B_{c2}$ to the dominating field-independent
bulk term  $\sim N_F T^2/T_c $. Two terms become comparable at the
temperature $  T/T_c\sim (B/B_{c2})^{1/2}$, where the crossover occurs to the
square-root behavior $C(T)\sim T N_F \sqrt{B/B_{c2}}$ of
Ref.\cite{d-waveVortex}  at lower temperatures, $  T/T_c\ll (B/B_{c2})^{1/2}
\ll 1$.

Both high-temperature and low-temperature  asymptotic dependences of the
specific heat on the magnetic field,  $ N_F T_c  B/B_{c2}$ and $   N_FT
\sqrt{B/B_{c2}}$,   come due to the velocity field far from the vortex and
do not depend on the details of the vortex core structure discussed in
\cite{Franz,Ichioka}.

This work was supported by the Russian
Foundation for Fundamental Sciences, Grant No. 96-02-16072.

\end{document}